\documentclass[twocolumn,showpacs,aps,pra,amsfonts,amsmath,amssymb,superscriptaddress,floatfix]
{revtex4}
\usepackage[dvips]{graphicx}
\usepackage{amsmath}
\usepackage{amssymb}

\newcommand{\beq}{\begin{equation}}
\newcommand{\eeq}{\end{equation}}

\begin{document}

\title{Large amplitude coherent state superposition generated by 
a time-separated two-photon subtraction from a continuous wave squeezed vacuum}

\author{Masahiro Takeoka}
\affiliation{
    National Institute of Information and Communications 
    Technology, 
    4-2-1 Nukui-Kita, Koganei, Tokyo 184-8795, Japan}
\affiliation{
    CREST, Japan Science and Technology Agency, 
    5 Sanbancho, Chiyoda-ku, Tokyo 102-0075, Japan}

\author{Hiroki Takahashi}
\affiliation{
    National Institute of Information and Communications 
    Technology, 
    4-2-1 Nukui-Kita, Koganei, Tokyo 184-8795, Japan}
\affiliation{
    CREST, Japan Science and Technology Agency, 
    5 Sanbancho, Chiyoda-ku, Tokyo 102-0075, Japan}
\affiliation{
    Department of Applied Physics, The University of Tokyo 
    7-3-1 Hongo, Bunkyo-ku, Tokyo 113-8656, Japan}

\author{Masahide Sasaki}
\affiliation{
    National Institute of Information and Communications 
    Technology, 
    4-2-1 Nukui-Kita, Koganei, Tokyo 184-8795, Japan}
\affiliation{
    CREST, Japan Science and Technology Agency, 
    5 Sanbancho, Chiyoda-ku, Tokyo 102-0075, Japan}

\date{\today}

\begin{abstract}

Theoretical analysis is given for a two-photon subtraction 
from a continuous wave (cw) squeezed vacuum with finite time separation 
between two detection events. 
In the cw photon subtraction process, generated states are inevitably 
described by temporal multimode states. 
Our approach is based on analytical formulae 
that are mathematically simple and provide 
an intuitive understanding of the multimode structure of the system. 
We show that, in our process, the photon subtracted squeezed vacuum is 
generated in two temporal modes and one of these modes acts as 
an ancillary mode to make the other one a large amplitude coherent state 
superposition. 

\end{abstract}
\pacs{03.67.Hk, 42.50.Dv}

\maketitle

\section{Introduction
\label{sec:intro}}

Photon subtraction is a useful technique to conditionally manipulate 
nonclassical state of light.  
To subtract photons from a traveling wave, one puts 
a highly transmissive beam splitter (BS) and detects 
a reflected beam by photo-detectors.  
Selection of the event such that the detectors observe $n$ photons in total 
approximately acts as annihilation operations $\hat{a}^n$ 
to the initial state. 
Applications of the photon subtraction have been proposed so far, e.g. 
generation of Schr\"{o}dinger cat-like states \cite{Dakna97}, 
distillation of Gaussian entangled states \cite{Browne03}, 
and loophole free tests of the violation of Bell's inequality 
\cite{Nha04,G-Patron04}. 
In addition, by adding 
displacement operations in front of the detectors, one would obtain 
higher flexibility of the output state synthesis 
\cite{Fiurasek05,Takeoka07,Nielsen07-1}.

Here the optical Schr\"{o}dinger cat state means the superposition of 
distinct coherent states 
\begin{equation}
\label{eq:CatState}
|C_\pm \rangle = \frac{1}{\sqrt{\mathcal{N}_\pm}} 
\left( |\alpha\rangle \pm |-\alpha\rangle \right) ,
\end{equation}
and an $n$-photon subtracted squeezed vacuum with even (odd) $n$ 
is approximately equal to $|C_+\rangle$ ($|C_-\rangle$). 
Recently, single-photon subtraction from a squeezed vacuum has 
been experimentally demonstrated with a pulsed \cite{Ourjoumtsev06} 
and cw \cite{N-Nielsen06,Wakui07} squeezed vacuum, 
those correspond to the generation of 
$|C_-\rangle$ with $|\alpha|^2 \approx 1$. 
It should also be noted that an alternative way by using a photon number 
state and conditional homodyne detection has been proposed and experimentally 
demonstrated with a pulsed two-photon state \cite{Ourjoumtsev07}.  
The conditional output corresponds to a superposition of 
displaced $X$-squeezed states $\hat{D}(\pm\beta)S(r)|0\rangle$ 
with $|\beta|^2 \approx 1.2$ which would become $|C_+\rangle$ 
with $|\alpha|^2 \approx 2.6$ after applying appropriate 
squeezing operation in $P$-direction. 
Once such a larger amplitude coherent state superposition becomes 
directly available, 
it would be an important resource for the linear optics quantum computation 
scheme \cite{Ralph03}.

For cw experiments, one has to take into account a problem of 
mode mismatch between a squeezed state and photon detection events. 
While coherence time of the cw squeezed vacuum generated from 
an optical parametric oscillator (OPO) 
is given by the inverse of the OPO cavity bandwidth ($\zeta_0^{-1}$), 
the photon detection usually occurs within almost instantaneous 
time duration ($\Delta\tau \ll \zeta_0^{-1}$). 
As a consequence, the conditional output state appears locally around 
the photon detection time with a nontrivial mode function. 
General theory to simulate such experiments has been developed 
in \cite{Sasaki06,Molmer06} and the optimal mode functions are investigated 
in detail for the conditional generation of single- and 
two-photon states from a two-mode squeezed vacuum (non-degenerate OPO) 
\cite{Nielsen07single,Nielsen07multi}.

In this paper, we theoretically investigate a two-photon subtraction 
from a cw squeezed vacuum generated by a degenerate OPO. 
For a pulsed scheme (or the original proposal in \cite{Dakna97}), 
it conditionally generates an even parity superposition 
$|C_+\rangle$ with $|\alpha|^2 \approx 1$. 
Compared to the pulsed scheme, a distinct feature of the cw scheme 
is that the photon detection events generally occur in different times. 
We show that, surprisingly, with appropriate time difference $\Delta$, 
a size of the cat state is drastically increased 
to $|\alpha|^2 = 2.5 \sim 3$ with the fidelity of $F > 0.9$. 
It is shown that the time-separated photon subtraction generates a nontrivial 
two-mode state distributed in two particular temporal modes 
which allows us to synthesize the output state and 
results a larger amplitude coherent state superposition 
in an appropriate temporal mode. 
Also, contrasted to the previous theoretical analyses of 
the cw schemes \cite{Sasaki06,Molmer06,Nielsen07single,Nielsen07multi}, 
our approach is mostly analytical, 
which is useful for intuitive understandings of the multimode structure 
in the generated state and how the size of the cat state is increased 
via quantum interferences.

The paper is organized as follows. 
In Sec.~II, before treating cw sources, we briefly review the properties 
of the photon subtracted squeezed state and coherent state superposition. 
In Sec.~III, we discuss the cw time-separated photon subtraction in detail and 
show how we can extract a larger coherent state superposition from 
a temporal multimode photon subtracted state. 
We also discuss the physical insight of the same scheme 
from an alternative point of view 
in Sec.~IV, where we show that the two types of quantum interferences 
play crucial roles to increase the size of the superposed coherent states. 
Section V concludes the paper.

\section{Synthesis of approximate coherent state superpositions
\label{sec:cat_and_ps}}

Generation of a Schr\"{o}dinger cat-like state by 
photon subtraction from a traveling wave squeezed vacuum was proposed 
in \cite{Dakna97} which consists of 
a small reflectance beam splitter and, in ideal, a photon number resolving 
detector at the reflected port. 
When the detector counts $n$ photons, the transmitted state is 
transformed to $|nPS\rangle \propto \hat{a}^n\hat{S}(-r)|0\rangle$, 
which approximates $|C_+\rangle$ ($|C_-\rangle$) for even (odd) $n$, 
where $\hat{S}(r) = \exp[\frac{r}{2}(\hat{a}^2 - \hat{a}^{\dagger \, 2})]$ 
is the squeezing operator and $-r$ represents the squeezing in 
$P$ direction of the phase space. 
Asymptotically, for large $n$, the size of $\alpha$ increases 
as $|\alpha|^2 = n$.

Since $\hat{S}(r)^\dagger \hat{a} \hat{S}(r) = 
\hat{a} \cosh r - \hat{a}^\dagger \sinh r$, 
$n$-photon subtracted state $\hat{a}^n\hat{S}(-r)|0\rangle$ is 
rewritten as a squeezed state of a superposition of 
even or odd number states as 
\begin{eqnarray}
\label{eq:single-mode-Bogolubov2}
|nPS\rangle & = & \frac{1}{\sqrt{\mathcal{N}_n}} 
\hat{a}^n \hat{S}(-r) |0\rangle 
\nonumber\\ & = & 
\frac{1}{\sqrt{\mathcal{N}_n}} 
\hat{S}(-r) ( \hat{a} \cosh r + \hat{a}^\dagger \sinh r )^n |0\rangle 
\nonumber\\ & = & 
\hat{S}(-r) \left\{ 
\begin{array}{l}
( c_{n}^r |n\rangle + c_{n-2}^r |n-2\rangle 
+ \cdots + c_0^r |0\rangle ) 
\\ \quad 
n: {\rm even} , \\
( c_{n}^r |n\rangle + c_{n-2}^r |n-2\rangle 
+ \cdots + c_1^r |1\rangle ) 
\\ \quad 
n: {\rm odd} , 
\end{array}
\right.
\nonumber\\
\end{eqnarray}
where $\mathcal{N}_n$ is a normalization factor and 
each of $c_i^r$ is a function of $r$. 
For example, a single-photon subtracted state is exactly 
equivalent to a squeezed single-photon state.

Meanwhile, it has recently been predicted that 
if one could arbitrarily synthesize the superposition ratio $c_i$ 
in the last line of Eq.~(\ref{eq:single-mode-Bogolubov2}), 
approximate $|C_\pm\rangle$ with larger $\alpha$ would be obtainable 
\cite{Ourjoumtsev07,Nielsen07-1}. 
To discuss it more precisely, let us consider the state 
\begin{equation}
\label{eq:psi_single_mode}
|\psi_n\rangle = \hat{S}(-r)(c_n |n\rangle + c_{n-2} |n-2\rangle \cdots ) ,
\end{equation}
and suppose we can arbitrarily set $c_i$'s. 
Since the fidelity between $|C_\pm\rangle$ and $|\psi_n\rangle$ 
is described as 
\begin{equation}
\label{eq:fidelity_single_mode}
F=|\langle C_\pm|\hat{S}(-r)(c_n |n\rangle + \cdots )|^2 ,
\end{equation}
the optimal coefficients $c_i$'s maximizing $F$ are proportional 
to those of the $X$-squeezed $|C_\pm\rangle$ 
\begin{equation}
\label{eq:squeezed_cat}
\hat{S}(r) |C_\pm\rangle = 
\frac{1}{\mathcal{N}_\pm} \hat{S}(r) \left( 
 |\alpha\rangle \pm |-\alpha\rangle \right) ,
\end{equation}
up to $n$. 
Here 
\begin{eqnarray}
\label{eq:squeezed_coh}
\hat{S}(r) |\pm\alpha\rangle & = & 
(1-\lambda^2)^{1/4} e^{-(1-\lambda)\alpha^2/2}
\sum_{m=0}^\infty \frac{1}{\sqrt{m!}} 
\nonumber\\ && \times 
\left( \frac{\lambda}{2} \right)^{n/2} 
H_m \left( \pm\sqrt{\frac{1-\lambda^2}{2\lambda}} \alpha \right) 
|m\rangle , 
\end{eqnarray}
where $\lambda = \tanh r$ and $H_m (x)$ is a Hermite polynomial.

\begin{figure}
\begin{center}
\includegraphics[width=0.8\linewidth]{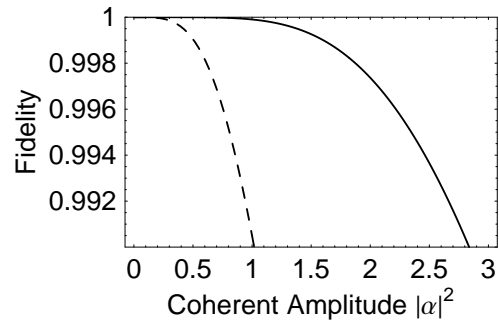}   %
\caption{\label{fig:alpha-fidelity}
Fidelity vs the size of the cat state $\alpha$ for  
$F=|\langle C_+|\psi_2\rangle|^2$ (solid line) and 
$F=|\langle C_+|2PS\rangle|^2$ (dashed line). 
}
\end{center}
\end{figure}

For example, when $n=2$, the two-photon subtracted state is given by 
\begin{eqnarray}
\label{eq:tps}
|2PS\rangle & \propto & 
\hat{a}^2 \hat{S}(-r) |0\rangle 
\nonumber\\ & = & \hat{S}(-r) 
\sinh r \left( \sqrt{2} \sinh r |2\rangle + \cosh r |0\rangle \right).
\end{eqnarray}
The squeezed superposition state $|\psi_2\rangle$ is on the other hand found 
from Eq.~(\ref{eq:squeezed_cat}) to be 
\begin{eqnarray}
\label{eq:tps}
|\psi_2\rangle & \propto & 
\hat{S}(-r) \left( \frac{(1-\lambda^2)\alpha^2 - \lambda}{\sqrt{2}} |2\rangle 
+ |0\rangle \right) .
\end{eqnarray}
The fidelities between these states and $|C_+\rangle$ are 
plotted in Fig.~(\ref{fig:alpha-fidelity}) where note that, 
for given $\alpha$, the squeezing parameter $r$ is optimized 
to maximize the fidelities. 
It is clearly shown that one can obtain $|C_+\rangle$ 
with more than twice larger average power ($|\alpha|^2 = 2.5 \sim 3$) 
if it is possible to synthesize the number state superposition.

In the next section, we show that, with a cw squeezed vacuum source, 
one can synthesize the superposition by simply having a time separation 
between two photo-detection events.

\section{Time-separated photon subtraction from a cw squeezed vacuum}

\begin{figure}
\begin{center}
\includegraphics[width=1\linewidth]{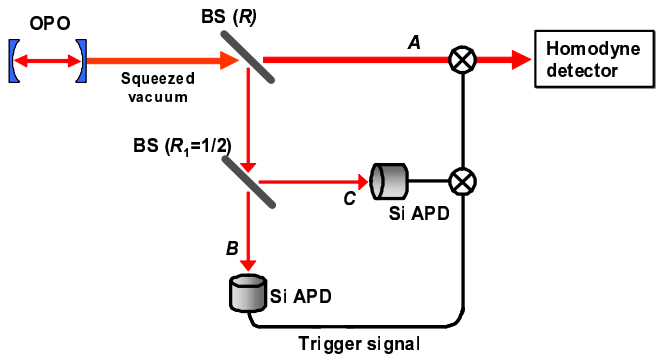}   %
\caption{\label{fig:schematic1}
(Color online) Schematic of the two-photon subtraction from 
a cw squeezed vacuum.
}
\end{center}
\end{figure}

In this section, we derive an analytical expression of 
the time-separated two-photon subtracted squeezed vacuum. 
Schematic is shown in Fig.~\ref{fig:schematic1}. 
A cw squeezed vacuum generated from an OPO with the bandwidth $\zeta_0$ 
is split via a small reflectance BS with the reflectance $R$ and 
the reflected fraction is guided into two photon detectors in path B and C 
through a half BS ($R_1=1/2$). 
Choosing the event such that two photons click each of detectors at time 
$t=t_1$ and $t_2$, where $|t_2-t_1|$ is comparable 
or smaller than $\zeta_0^{-1}$, one can conditionally obtain a two-photon 
subtracted state as an output in path A, which is measured 
by a homodyne detector.

For cw sources, it is important to take into account the mode mismatch 
between the squeezed vacuum and the photon detection events 
since the squeezed vacuum is a multi-mode state distributed within 
a time $\zeta_0^{-1}$, 
while the photon detections happen almost instantaneously 
\cite{Sasaki06,Molmer06,Nielsen07single,Nielsen07multi}. 
The conditional output we want to observe is therefore generated 
in a temporal mode localized around $t_1$ and $t_2$. 
To detect such state, the homodyne measurement might be  
a time integrating detection with an appropriate mode function filter. 
In other words, the homodyne measurement extracts one particular temporal 
mode from the cw signal.

In the following subsections, we first give a slight modification of 
the usual input-output theory of the degenerate OPO, and then 
describe the cw photon subtraction process in Schr\"{o}dinger picture. 
Since the purpose of this section is to describe the structure of 
the time-separated photon subtracted state, we assume that all elements 
in the scheme are lossless and there is no technical noise. 
We show that, in an appropriately filtered conditional state, 
the number state superposition is well synthesized 
via the time difference of photo-detection events.

\subsection{Squeezing operation via an ideal degenerate OPO}

Let us start by briefly reviewing the input-output theory 
of the degenerate OPO \cite{Collet84}. 
Denote the positive-frequency part of the field operator 
in a single transverse mode of the optical field 
with a continuous spectrum by 
\beq  
\hat a(t)=\frac{1}{2\pi}\int_{-\infty}^{\infty} d\Omega
\hat a(\omega_0+\Omega)
e^{-i(\omega_0+\Omega)t}. 
\eeq
Here $\hat{a}(\omega_0+\Omega)$ is the annihilation operator 
for the Fourier component at angular frequency $\omega_0+\Omega$, 
where 
$\omega_0$ is the center angular frequency of the OPO. 
The operator obeys the continuum commutation relation 
\beq\label{continuum commutation relation in frequency} 
[\hat a(\omega_0+\Omega), \hat a^\dagger(\omega_0+\Omega')]
=2\pi\delta(\Omega-\Omega'). 
\eeq
The time dependent field operator $\hat a(t)$ is defined in 
the interval $(-\infty,\infty)$, 
and obeys the commutation relation 
\beq\label{continuum commutation relation in time} 
[\hat a(t), \hat a^\dagger(t')]=\delta(t-t'). 
\eeq

These operators should be moved to 
a rotating frame at 
the center frequency $\omega_0$, 
\beq\label{operator in rotating frame}
\hat A(t)=\hat a(t) e^{i\omega_0t}
=
\frac{1}{2\pi}\int_{-\infty}^{\infty} d\Omega
\hat A(\Omega) e^{-i\Omega t}, 
\eeq
where 
$\hat A(\Omega)=\hat a(\omega_0+\Omega)$.

Following \cite{Collet84}, 
an input-output relation of the lossless OPO is described 
by the Bogolubov transformation 
\begin{equation}
\label{eq:Lossless_Bogolubov}
\hat{S}_A^\dagger \hat{A} (\Omega) \hat{S}_A = 
\mu(\Omega) \hat{A}(\Omega) 
+ \nu(\Omega) \hat{A}^\dagger(-\Omega) ,
\end{equation}
where 
\begin{eqnarray}
\label{eq:Lossless_mu_nu}
\mu(\Omega) & = & 
\frac{\zeta_0^2 + \epsilon^2 + \Omega^2}{
(\zeta_0 - i\Omega)^2 - \epsilon^2} ,
\\
\nu(\Omega) & = & 
\frac{2 \zeta_0 \epsilon}{(\zeta_0 - i\Omega)^2 - \epsilon^2} .
\end{eqnarray}
Here $\zeta_0 \equiv \gamma_T/2$ corresponds to the bandwidth of 
the OPO output, and $\gamma_T$ and $\epsilon$ are 
the leakage rate of the OPO output mirror and 
the parameter proportional to the nonlinear coefficient of the OPO crystal 
and the pump amplitude, respectively.

The above transformation is rewritten in a useful form 
by introducing the input field annihilation operator
\begin{eqnarray}
\label{eq:A_bar}
\hat{A}_\theta(\Omega) & \equiv & \hat{A} (\Omega) e^{i\theta(\Omega)} ,
\end{eqnarray}
with 
\begin{equation}
\label{eq:theta_omega}
\tan\theta(\Omega) = 
\frac{2 \zeta_0 \Omega}{\zeta_0^2 - \epsilon^2 - \Omega^2} ,
\end{equation}
which obeys the communication relation 
\begin{equation}
\label{eq:A_commutation_relation}
[ \hat{A}_\theta(\Omega) , \hat{A}_\theta^\dagger(\Omega')]
=2\pi \delta(\Omega-\Omega') .
\end{equation}
The Bogolubov transformation in Eq.~(\ref{eq:Lossless_Bogolubov}) 
is then rewritten as 
\begin{equation}
\label{eq:Lossless_Bogolubov2}
\hat{S}_A^\dagger \hat{A} (\Omega) \hat{S}_A = 
\bar{\mu}(\Omega) \hat{A}_\theta (\Omega) 
+ \bar{\nu}(\Omega) 
\hat{A}_\theta^\dagger(-\Omega) ,
\end{equation}
where
\begin{eqnarray}
\label{eq:Lossless_mu_nu_bar}
\bar{\mu}(\Omega) & \equiv & \mu(\Omega) e^{-i\theta(\Omega)} = 
\frac{\zeta_0^2 + \epsilon^2 + \Omega^2}{
\sqrt{(\zeta_+^2 + \Omega^2)
(\zeta_-^2 + \Omega^2)}} ,
\\
\bar{\nu}(\Omega) & \equiv & \nu(\Omega) e^{-i\theta(\Omega)} = 
\frac{2 \zeta_0 \epsilon}{
\sqrt{(\zeta_+^2 + \Omega^2) 
(\zeta_-^2 + \Omega^2)}} ,
\end{eqnarray}
and $\zeta_\pm = \zeta_0 \pm \epsilon$.

\subsection{Schr\"{o}dinger picture of the two-photon subtraction 
from a cw squeezed vacuum}

Let us now turn to the photon subtraction operations. 
Denote the time dependent field operator in paths A, B, and C 
(in a rotating frame) by $\hat{A}(t)$, $\hat{B}(t)$, and $\hat{C}(t)$
, respectively (see Fig.~~\ref{fig:schematic1}). 
A cw squeezed vacuum state generated from a lossless OPO is described by 
$\hat{S}_A |{\bf 0}_A\rangle$. 
After beam splitting it via the two BSs ($R$ and $R_1$), 
the quantum state distributed in three paths 
is described by $\hat{V}_{BC} \hat{V}_{AB} \hat{S}_A |{\bf 0}_{ABC}\rangle$ 
where $\hat{V}_{AB}$ is a beam splitting operator transforming 
the field operator, for example,  
$\hat{V}_{AB}^\dagger \hat{B}(t) \hat{V}_{AB} 
= - \sqrt{R} \hat{A}(t) + \sqrt{1-R} \hat{B}(t)$. 
Suppose the first and second photons are detected at the time $t=t_1$ and 
$t_2$ in paths B and C, respectively, while the detectors project the state 
onto vacua in all other time. 
We assume that the detector's time resolution is instantaneous i.e. 
enough shorter than the cavity lifetime $\zeta_0^{-1}$. 
After such event, the conditional output state in path A is projected onto 
\begin{eqnarray}
\label{eq:TPS-pre}
|\rho_{cw} \rangle & \propto & 
\langle {\bf 0}_{BC}| \hat{C}(t_2) \hat{B}(t_1) 
\hat{V}_{BC} \hat{V}_{AB} \hat{S}_A |{\bf 0}_{ABC}\rangle 
\nonumber\\ & = & 
\frac{R}{\sqrt{2}} \exp\left[ 
\ln \sqrt{1-R} 
\int_{-\infty}^\infty dt \, \hat{A}^\dagger (t) \hat{A} (t) \right] 
\nonumber\\ && \times 
\hat{A}(t_2) \hat{A}(t_1) \hat{S}_A | {\bf 0}_A \rangle 
\end{eqnarray}
where note that $R_1=1/2$. 
In the limit of small $R$, the exponential term is approximated to be 
$\exp [ \ln \sqrt{1-R} \int dt \hat{A}^\dagger (t) \hat{A} (t)]
\sim 1$ and thus the state is described by a two-photon 
annihilated squeezed vacuum 
\begin{equation}
|\rho_{cw}\rangle \propto 
\hat{A}(t_2) \hat{A}(t_1) \hat{S}_A | {\bf 0}_A \rangle .
\end{equation}
By use of the property of the squeezing operation 
in Eq.~(\ref{eq:Lossless_Bogolubov2}), we further obtain 
\begin{eqnarray}
\label{eq:TPS}
|\rho_{cw} \rangle & \propto & 
\hat{A}(t_2) \hat{A}(t_1) \hat{S}_A | {\bf 0}_A \rangle 
\nonumber\\ & = & 
\hat{S}_A \left( 
\int_{-\infty}^\infty dt \bar{\nu}(t-t_2) \hat{A}_\theta^\dagger(t) 
\int_{-\infty}^\infty dt \bar{\nu}(t-t_1) \hat{A}_\theta^\dagger(t) 
\right. \nonumber\\ && \left. 
+ \int_{-\infty}^\infty dt 
\bar{\mu}(t-t_2) \bar{\nu}(t-t_1) 
\right) |{\bf 0}_A \rangle ,
\end{eqnarray}
where 
\begin{eqnarray}
\label{eq:mu_nu_bar_FT}
\bar{\mu}(t-t_0) & = & \frac{1}{2\pi} 
\int_{-\infty}^\infty d\Omega \bar{\mu}(\Omega) e^{-i\Omega(t-t_0)} ,
\\
\bar{\nu}(t-t_0) & = & \frac{1}{2\pi} 
\int_{-\infty}^\infty d\Omega \bar{\nu}(\Omega) e^{-i\Omega(t-t_0)} .
\end{eqnarray}
Defining the normalized temporal function 
\begin{eqnarray}
\label{eq:Psi_0_def}
\psi (t-t_0) & \equiv & \frac{1}{\sqrt{{\mathcal N}_\nu}} \bar{\nu}(t-t_0) ,
\\ 
\label{eq:N_nu}
{\mathcal N}_\nu & = & \frac{\zeta_0 \epsilon}{2} 
\left( \frac{1}{\zeta_-} - \frac{1}{\zeta_+} \right) ,
\end{eqnarray}
the output state is simply described as 
\begin{eqnarray}
\label{eq:TPS_2}
|\rho_{cw}\rangle & \propto & 
 \hat{S}_A \left( {\mathcal N}_\nu \hat{A}_2^\dagger \hat{A}_1^\dagger
+ F_\Delta \right) |{\bf 0}_A \rangle ,
\end{eqnarray}
where
\begin{eqnarray}
\label{eq:A^dagger_i}
\hat{A}_i^\dagger & \equiv & 
\int_{-\infty}^\infty dt 
\psi(t-t_i) \hat{A}_\theta^\dagger (t),
\end{eqnarray}
\begin{eqnarray}
\label{eq:F_delta}
F_\Delta & = & 
\int_{-\infty}^\infty dt 
\bar{\mu}(t-t_2) \bar{\nu}(t-t_1) ,
\end{eqnarray}
and $\Delta = |t_2 - t_1|$ is the time difference between 
the photon detection events. 
Note that, in the limit of small $\epsilon$, the temporal mode function 
is approximated to be $\psi (t) \approx \sqrt{\zeta_0} e^{-\zeta_0 |t|}$ 
which is a reasonable approximation in the realistic experiments 
\cite{N-Nielsen06,Wakui07}. 
Now, since $\psi (t-t_i)$'s are nonorthogonal to each other, 
it is useful to introduce an orthonormal basis in these two-dimensional 
temporal modes. 
Let us here recall that we will observe only a single-temporal mode 
at the final homodyne detection step, which should be 
a highly nontrivial state.  
One of the natural choices of the such mode is to include 
the first term in the right hand side of Eq.~(\ref{eq:TPS_2}) 
as much as possible. 
Then we arrive at the basis consisting of symmetric and anti-symmetric 
orthonormal functions defined by 
\begin{equation}
\label{eq:orthonormal_mode}
\Psi_\pm (t) \equiv \frac{\psi (t-t_2) \pm \psi (t-t_1)}{
\sqrt{2 ( 1 \pm I_\Delta ) }} ,
\end{equation}
where 
\begin{eqnarray}
\label{eq:I_Delta}
I_\Delta & = & \int_{-\infty}^\infty dt \, \psi (t-t_2) \psi (t-t_1) 
\end{eqnarray}
and $\Psi_+(t)$ is expected to be the final mode, i.e. the filter function 
of the homodyne detection. 
Corresponding field operators and state vectors are also defined as 
\begin{equation}
\label{eq:orthonormal_mode}
\hat{A}_\pm^\dagger \equiv \int_{-\infty}^\infty dt 
\Psi_\pm (t) \hat{A}_\theta^\dagger (t) ,
\end{equation}
and 
\begin{equation}
\label{eq:+-_state_vector}
\hat{A}_\pm^\dagger |n_\pm\rangle \equiv 
\sqrt{n+1}|n+1_\pm\rangle , \quad
|{\bf 0}_A\rangle \equiv |0_+\rangle|0_-\rangle|{\bf 0}_{\tilde{A}}\rangle .
\end{equation}
Then $|\rho_{cw}\rangle$ is now expressed as 
\begin{eqnarray}
\label{eq:TPS_pure_2}
|\rho_{cw}\rangle & = & 
\frac{\hat{S}_A}{\sqrt{\mathcal N}} \left\{ \mathcal{N}_\nu \left( 
\frac{1+I_\Delta}{2} \hat{A}_+^{\dagger \, 2} 
- \frac{1-I_\Delta}{2} \hat{A}_-^{\dagger \, 2} \right)
+ F_\Delta \right\} |{\bf 0}_A \rangle
\nonumber\\ & = & 
\frac{\hat{S}_A}{\sqrt{\mathcal N}} \left\{ 
\frac{{\mathcal N}_\nu (1+I_\Delta)}{\sqrt{2}} |2_+,0_-\rangle 
\right. \nonumber\\ && \left. 
- \frac{{\mathcal N}_\nu (1-I_\Delta)}{\sqrt{2}} |0_+,2_-\rangle 
+ F_\Delta |0_+,0_-\rangle \right\} |{\bf 0}_{\tilde{A}}\rangle ,
\nonumber\\ & = & 
\hat{S}_A |\rho_{+-}\rangle |{\bf 0}_{\tilde{A}}\rangle ,
\end{eqnarray}
where 
\begin{eqnarray}
\label{eq:Normalization}
{\mathcal N} & = & 
{\mathcal N}_\nu^2 (1+I_\Delta^2) + F_\Delta^2 .
\end{eqnarray}
Note that, in Eq.~(\ref{eq:TPS_pure_2}), 
the cross-term $\hat{A}_+^\dagger \hat{A}_-^\dagger$ is vanished 
due to the bunching-like interference of these creation operators. 
We will discuss such quantum interference again in the next section.

The conditional output state described by Eq.~(\ref{eq:TPS_pure_2}) 
is understood as the squeezed state of 
$|\rho_{+-}\rangle|{\bf 0}_{\tilde{A}}\rangle$ 
where $|\rho_{+-}\rangle$ is the superposition of 
the two-photon states occupying the temporal mode $\Psi_+(t)$ or 
$\Psi_-(t)$, respectively, and the vacuum state.
Note that when two photons are detected at the same time ($\Delta=0$), 
the second term of $|\rho_{+-}\rangle$ vanishes 
(mode $\Psi_-(t)$ cannot be defined) and thus 
the ideal two-photon subtracted state is localized into the temporal 
mode $\Psi_+(t)$.  
Moreover, as will be shown, when $\Delta$ increases, 
one can observe a larger amplitude coherent state superposition 
in mode $\Psi_+(t)$. 
In the following, therefore, we choose $\Psi_+(t)$ as the filter function 
for the temporal mode of the LO field in the homodyne detector, 
i.e. the detector's observable is described by the quadrature operator 
\begin{equation}
\label{eq:homodyne_observable}
\hat{X}_{HD} \equiv \int_{-\infty}^{\infty} dt \, \Psi_+(t) 
\hat{X}(t), 
\end{equation}
(in practice, the integration is carried within a finite time width 
$T'$ ($\gg \zeta_0^{-1}$)).

Let us see the reduced quantum state in mode $\Psi_+(t)$. 
We first trace out mode $\tilde{A}$ from $|\rho_{cw}\rangle$. 
The reduced state is then given by 
\begin{eqnarray}
\label{eq:rho_+-}
\hat{\rho}_{+-} & = & 
{\rm Tr}_{\tilde{A}}[|\rho_{cw}\rangle\langle\rho_{cw}|]
\nonumber\\ & = & 
\left(  \hat{\mathcal{S}}_+ 
\otimes \hat{\mathcal{S}}_- \right) 
|\rho_{+-}\rangle\langle\rho_{+-}| ,
\end{eqnarray}
where $\hat{\mathcal{S}}_+ $ and 
$\hat{\mathcal{S}}_- $ are the non-unitary Gaussian operations 
acting on modes $\Psi_+$ and $\Psi_-$ separately. 
Each of $\hat{\mathcal{S}}_\pm$ can be regarded as 
a single-mode squeezing operation although 
the process includes a small coupling with thermal environment 
(for derivation and characteristics of $\hat{\mathcal{S}}_\pm$, 
see Appendix A).

Let us further trace out mode $\Psi_-(t)$ from $\hat{\rho}_{+-}$. 
We then obtain 
\begin{eqnarray}
\label{eq:final_state_+}
\hat{\rho}_+ & = & {\rm Tr}_-[\hat{\rho}_{+-}] 
\nonumber\\ & = & 
C_\phi \hat{\mathcal S}_+ |\phi\rangle\langle\phi|
+ C_v  \hat{\mathcal S}_+ |0 \rangle\langle 0| ,
\end{eqnarray}
where 
\begin{eqnarray}
\label{eq:phi_cat}
|\phi\rangle & = & 
c_2 |2_+\rangle + c_0 |0_+\rangle 
\nonumber\\ & = & 
\frac{1}{\sqrt{{\mathcal N}_\phi}} 
\left\{ \frac{{\mathcal N}_\nu (1+I_\Delta)}{\sqrt{2}} |2_+\rangle 
+ F_\Delta |0_+\rangle \right\} ,
\end{eqnarray}
and 
\begin{eqnarray}
\label{eq:N_phi}
{\mathcal N}_\phi & = & 
\frac{{\mathcal N}_\nu}{2} (1+I_\Delta)^2 + F_\Delta^2 .
\end{eqnarray}
As shown in these equations, 
the output quantum state observed by the homodyne detector 
with the filter function $\Psi_+(t)$ 
is the statistical mixture of the squeezed 
$|\phi\rangle$ and vacuum. 
The ratio of the statistical mixing is given by 
\begin{equation}
\label{eq:C_phi}
C_\phi = \frac{{\mathcal N}_\phi}{\mathcal N} 
= \frac{\frac{1}{2}{\mathcal N}_\nu^2 (1+I_\Delta)^2 + F_\Delta^2}{
{\mathcal N}_\nu^2 (1+I_\Delta^2) + F_\Delta^2} ,
\end{equation}
and $C_v=1-C_\phi$.

\begin{figure}
\begin{center}
\includegraphics[width=1.0\linewidth]{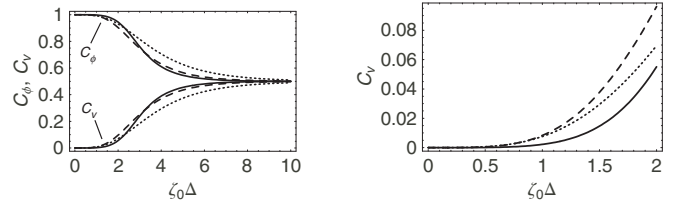}   %
\caption{\label{fig:cat_fraction}
The upper and lower lines show $C_\phi$ and $C_v$, respectively 
with $|\epsilon|/\zeta_0$=0.1 (solid lines), 0.3 (dashed lines), 
and 0.5 (dotted lines).
}
\end{center}
\end{figure}
\begin{figure}
\begin{center}
\includegraphics[width=1.0\linewidth]{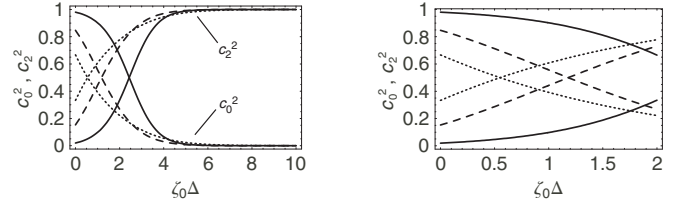}   %
\caption{\label{fig:c2-c0_fraction}
The upper and lower lines (at $\zeta_0 \Delta = 0$) show 
$c_0^2$ and $c_2^2$, respectively 
with $|\epsilon|/\zeta_0$=0.1 (solid lines), 0.3 (dashed lines), 
and 0.5 (dotted lines).
}
\end{center}
\end{figure}

In Figs.~\ref{fig:cat_fraction} and \ref{fig:c2-c0_fraction}, 
we plot $C_\phi$, $C_v$, $c_2^2$, and $c_0^2$, that characterize 
$\hat{\rho}_+$, as functions of $\Delta$. 
For large delay ($\Delta \gg 1/\zeta_0$), 
the output state goes to the mixture of two- and zero-photon state as 
\begin{equation}
\label{eq:2-0_mixture}
\hat{\rho}_+ \to \frac{1}{2} \hat{\mathcal S}_+ 
\Big( |2\rangle\langle2| + |0\rangle\langle0| \Big) ,
\end{equation}
which was already pointed out in the scheme of generating multi-photon 
state from a cw non-degenerate OPO source \cite{Nielsen07-1}.

In our scheme, on the other hand, the intermediate delay region 
($\Delta \sim \zeta_0^{-1}$) is particularly interesting. 
Since $C_v$ is almost negligible 
as shown in Fig.~\ref{fig:cat_fraction}, 
$\hat{\rho}_+$ can be approximated to be  
\begin{equation}
\label{eq:small_delta}
\hat{\rho}_+ \approx \hat{\mathcal S}_+ 
|\phi\rangle\langle\phi| .
\end{equation}
On the other hand, Fig.~\ref{fig:c2-c0_fraction} clearly shows that 
the tuning of $\Delta$ around $\zeta_0^{-1}$ provides us 
a wide controllability of the superposition ratio of 
two- and zero-photon states in $|\phi\rangle$. 
As discussed in Sec.~II, this allows us 
to generate larger coherent state superposition.

\begin{figure}
\begin{center}
\includegraphics[width=1.0\linewidth]{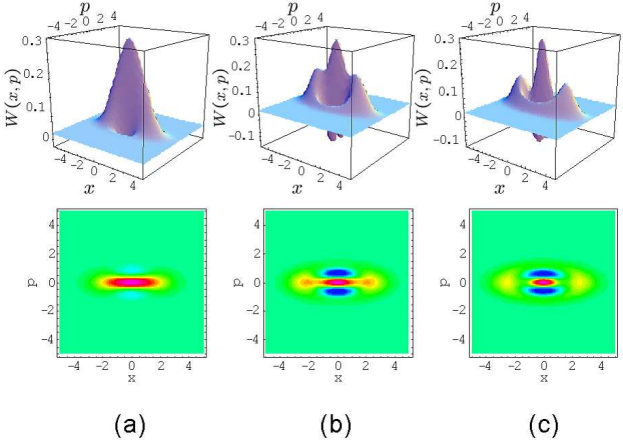}   %
\caption{\label{fig:rho_plus_Winger}
(Color online) 
Wigner functions of $\hat{\rho}_+$. $|\epsilon|/\zeta_0=0.27$ and 
(a) $\zeta_0 \Delta=0$, (b) $\zeta_0 \Delta=1.4$, (c) $\zeta_0 \Delta=2.4$. 
The fidelity between (b) and 
the ideal cat state with $|\alpha|^2=2.6$ is 0.946. 
}
\end{center}
\end{figure}

\begin{figure}
\begin{center}
\includegraphics[width=0.9\linewidth]{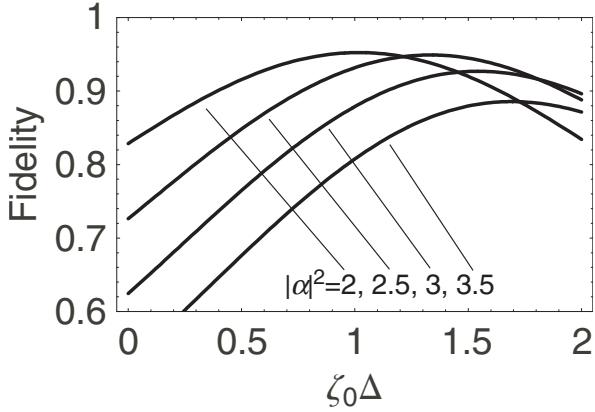}   %
\caption{\label{fig:rho_plus_fidelity}
Fidelity between $\hat{\rho}_+$ and the ideal cat state 
with $|\alpha|^2=$2, 2.5, 3, and 3.5 (from left to right). 
$|\epsilon|/\zeta_0 = 0.27$.
}
\end{center}
\end{figure}

To see the superposition property in phase space, 
we have calculated the Wigner function of $\hat{\rho}_+$ 
without any approximation (see Appendix B). 
Typical Wigner functions are shown 
in Figs.~\ref{fig:rho_plus_Winger}(a)-(c) for different $\Delta$. 
These figures clearly show that the increase of $\Delta$ induces 
a larger size superposition. 
Note that the fidelity between the state in Fig.~\ref{fig:rho_plus_Winger}(b) 
and $|C_+\rangle$ with $|\alpha|^2 = 2.6$ is 0.946.
Fidelities between $\hat{\rho}_+$ and $|C_+\rangle$'s with different 
amplitudes are shown in Fig.~\ref{fig:rho_plus_fidelity} 
as a function of $\Delta$. 
We see that our time-separated two-photon subtraction technique 
allows us to generate a larger superposition such as $|\alpha|^2 > 2.5$ 
with high fidelity ($0.9>$), that is not possible with a normal single-mode 
(pulsed) two-photon subtraction.


\section{Quantum interference of the subtracted photons}

In the previous section, we have observed a bosonic bunching-like effect 
between $\hat{A}^\dagger_+$ and $\hat{A}^\dagger_-$ that originates 
from the two-photon subtraction (see Eq.~(\ref{eq:TPS_pure_2})). 
In this section, we again discuss 
the cw two-photon subtraction from the viewpoint of quantum interferences. 
It is natural to expect that the state synthesis is achieved by 
some quantum interference. 
Generally, quantum interference occurs when the quantum process or 
states are going to be intrinsically indistinguishable. 
We concretely show that, in our case, 
two types of such interferences contribute to 
the size increase of the cat state.

Let us reexamine the cw state $|\rho_{cw}\rangle = 
\frac{1}{\sqrt{\mathcal{N}}} \hat{A}(t_2) \hat{A}(t_1) \hat{S}_A 
|{\bf 0}\rangle$ and consider from which mode the subtracted 
photons ($\hat{A}(t_2)$ and $\hat{A}(t_1)$) originates, 
$\Psi_+(t)$ or $\Psi_-(t)$. 
This is clarified by defining the following annihilation operators 
\begin{eqnarray}
\label{eq:a_+-}
\hat{A}_{t_\pm} & \equiv & 
\frac{1}{\sqrt{2}} \left( \hat{A}(t_2) \pm \hat{A}(t_1) \right) ,
\end{eqnarray}
where $\hat{A}_{t_+}$ ($\hat{A}_{t_-}$) is the annihilation 
operator which subtracts one photon from the squeezed vacuum 
in mode $\Psi_+(t)$ ($\Psi_-(t)$).

This is confirmed by directly substituting them as 
\begin{eqnarray}
\label{eq:psi_bunching}
|\rho_{cw}\rangle & = & 
\frac{1}{2\sqrt{\mathcal{N}}}  \left( \hat{A}_{t_+}^2 
- \hat{A}_{t_-}^2 \right) \hat{S}_A |{\bf 0}\rangle 
\nonumber\\ & = & 
\sigma_+ \hat{S}_A |\gamma_+\rangle |0_-\rangle |{\bf 0}_{\tilde{A}}\rangle 
- \sigma_- \hat{S}_A |0_+\rangle |\gamma_-\rangle |{\bf 0}_{\tilde{A}}\rangle ,
\nonumber\\
\end{eqnarray}
where we clearly see that $\hat{A}_{t_+}$ and $\hat{A}_{t_-}$ change 
the states only in $\Psi_+(t)$ and $\Psi_-(t)$, respectively.  
Moreover, from the first to second line, we observe a bunching-like 
quantum interference, i.e. the possibility such that 
each photon is subtracted from each of $\Psi_\pm(t)$ is vanished. 
Note that 
\begin{eqnarray}
\label{eq:phi_pm}
|\gamma_\pm\rangle & = & \frac{1}{\sigma_\pm} \left\{
\sqrt{2} \left( 
\frac{1 \pm e^{-\zeta_- \Delta}}{\zeta_-} 
- \frac{1 \pm e^{-\zeta_+ \Delta}}{\zeta_+}
\right) |2_\pm \rangle 
\right. \nonumber\\ && \left. 
+ 
\left( 
\frac{1 \pm e^{-\zeta_- \Delta}}{\zeta_-} 
+ \frac{1 \pm e^{-\zeta_+ \Delta}}{\zeta_+}
\right) |0_\pm \rangle \right\},
\end{eqnarray}
and 
\begin{eqnarray}
\label{eq:sigma_pm}
\sigma_\pm^2 & = & 
2 \left( 
\frac{1 \pm e^{-\zeta_- \Delta}}{\zeta_-} 
- \frac{1 \pm e^{-\zeta_+ \Delta}}{\zeta_+}
\right)^2 
\nonumber\\ && 
+ \left( 
\frac{1 \pm e^{-\zeta_- \Delta}}{\zeta_-} 
+ \frac{1 \pm e^{-\zeta_+ \Delta}}{\zeta_+}
\right)^2 . 
\end{eqnarray}

The final state to be measured is obtained by tracing modes 
$\tilde{A}$ and $\Psi_-(t)$ out. 
If $|0_-\rangle$ and $|\gamma_-\rangle$ in Eq.~(\ref{eq:psi_bunching}) 
are orthogonal (distinguishable), 
the final state is reduced to be a statistical mixture 
of $|0_+\rangle$ and $|\gamma_+\rangle$. 
These terms are, however, nonorthogonal, i.e. partially indistinguishable. 
One therefore may expect that some of the quantum coherence
between $|0_+\rangle$ and $|\gamma_+\rangle$ remains. 
This is the second type of quantum interference. 
To see this clearer, let us first simplify the equation by 
taking the approximation of the small squeezing limit. 
Neglecting the higher order of $\epsilon/\zeta_0$, we have
\begin{eqnarray}
\label{eq:psi_small_epsilon}
|\rho_{cw}\rangle & \approx & 
\sigma_+ \hat{S}_A \left( 
\frac{\epsilon \delta_+}{\zeta_0} \hat{A}_+^{\dagger \, 2} + 1 
\right) 
|0_+\rangle |0_-\rangle |{\bf 0}_{\tilde{A}}\rangle 
\nonumber\\ && 
- 
\sigma_- \hat{S}_A \left( 
\frac{\epsilon \delta_-}{\zeta_0} \hat{A}_-^{\dagger \, 2} + 1 
\right) 
|0_+\rangle |0_-\rangle |{\bf 0}_{\tilde{A}}\rangle .
\end{eqnarray}
where 
\begin{eqnarray}
\label{eq:sigma_small_epsilon}
\sigma_\pm & \approx & \frac{1 \pm e^{-\zeta_0\Delta}}{2} ,
\\
\label{eq:delta_small_epsilon}
\delta_\pm & = & 1 \pm \frac{\zeta_0\Delta e^{-\zeta_0\Delta}}{
1 \pm e^{-\zeta_0\Delta}} .
\end{eqnarray}
Note that $\delta_+$ is around $1 \sim 1.25$ and 
in practice, $\epsilon/\zeta_0$ is often regarded as 
an effective pumping parameter normalized by the OPO threshold. 
We should argue that the state where the two photons 
are subtracted at the same time $t_1$ ($\Delta=0$) is also given by 
\begin{equation}
\label{eq:degenerate_TPS_small_epsilon}
\frac{1}{\sqrt{\mathcal{N}}}\hat{A}^2(t_1) \hat{S}|{\bf 0}\rangle 
\approx \hat{S} \left[ \frac{\epsilon}{\zeta_0} 
\hat{A}_1^{\dagger \, 2} + 1 \right] |{\bf 0}\rangle , 
\end{equation}
which corresponds to the original photon subtraction scheme 
proposed in \cite{Dakna97} and here we call it 
a time-degenerate two-photon subtracted state.

Let us trace modes $\tilde{A}$ and $\Psi_-(t)$ out and 
see the reduced state with finite $\Delta$. 
Approximating the squeezing operation acting on modes $\Psi_+(t)$ 
and $\Psi_-(t)$ 
to be a separable unitary squeezing operator $\hat{S}_+ \otimes \hat{S}_-$ 
(see Appendix A), 
we obtain the reduced state as 
\begin{eqnarray}
\hat{\rho}_+ \approx ( 1 - C_v ) |\Phi\rangle\langle\Phi| 
+ C_v \hat{S}_+ |0\rangle\langle0| \hat{S}_+^\dagger ,
\end{eqnarray}
where $C_v = \frac{\epsilon^2}{2\zeta_0^2} (1-I_\Delta)^2 / 
\left\{ \frac{\epsilon^2}{\zeta_0^2} (1+I_\Delta^2) + e^{-2\zeta_0\Delta} 
\right\}$ and 
\begin{eqnarray}
|\Phi\rangle \propto 
\hat{S}_+ \left( \frac{\epsilon \delta_+}{\zeta_0} \hat{A}_+^{\dagger \, 2} 
+ 1 \right) |0\rangle - 
\tanh \left(\frac{\zeta_0\Delta}{2}\right)
\hat{S}_+ |0\rangle ,
\end{eqnarray}
these correspond to Eqs.~(\ref{eq:final_state_+}) and (\ref{eq:phi_cat}). 
Therefore, the component of the cat-like sate $|\Phi\rangle$ 
produced by the time-separated two-photon subtraction 
stems from the quantum interference (superposition) of 
the time-degenerate two-photon subtracted state and a squeezed 
vacuum. 
That is, the engineering of the state indistinguishability in 
mode $\Psi_-(t)$ allows us to synthesize the superposition ratio 
in mode $\Psi_+(t)$, in other words, provides 
the size controllability of $|\Phi\rangle$ in mode $\Psi_+(t)$. 
Note that, on the other hand, $|0_-\rangle$ and $|\gamma_-\rangle$ 
are only {\it partially} indistinguishable. 
Their distinguishable part creates the statistical mixture term 
$\hat{S}_+ |0\rangle\langle0| \hat{S}_+^\dagger$ with the factor of $C_v$.

\section{Conclusion}

In conclusion, we have theoretically investigated the time-separated
two-photon subtraction from a continuous wave squeezed vacuum. 
In a single-mode theory, a two-photon subtracted squeezed vacuum 
is regarded as a squeezed state of $c_2^r |2\rangle + c_0^r |0\rangle$ 
in which, however, the superposition coefficients $c_2^r$ and $c_0^r$ 
are not optimal to maximize its fidelity to a coherent state superposition.

In case of a cw scheme, one needs a multi-mode theory and 
we showed that 
when the time difference of the two photo-detection events is finite 
but within the coherence time of the squeezed vacuum, 
the conditional output is appeared within two temporal modes, 
as a squeezed 
superposition of $|20\rangle$, $|02\rangle$, and $|00\rangle$. 
Such superposition and a careful choice of the single mode 
function $\Psi_+(t)$ allow us to synthesize the even photon number 
superposition of the conditional output state and with appropriate 
parameters it results a generation of cat-like states 
which have more than 90\% fidelity with the coherent state superposition 
of $|\alpha|^2 >2.5$. 
We have also discussed the same issue 
from the viewpoint of quantum interference, 
which reveals 
how the conditional output state of the time-separated photon 
subtraction is deviated from that of the time-degenerate one, 
due to the quantum interferences. 
Our theoretical approach provides analytical expressions of the states, 
which would be further useful to investigate an intuitive physical picture 
of more complicated multi-mode cw quantum states.

\appendix
\section{Squeezing operation on temporal modes $\Psi_\pm(t)$ 
via an optical parametric oscillator}

In this Appendix, we discuss the input-output relation of the OPO 
from the input state prepared in modes $\Psi_\pm(t)$ 
to the output state in modes $\Psi_\pm(t)$. 
Namely, we look at the completely positive trace preserving map 
\begin{equation}
\hat{\mathcal{S}} \, \hat{\rho}_{+-} 
= {\rm Tr}_{\tilde{A}} \left[ \hat{S}_A \left(
\hat{\rho}_{+-} \otimes 
|{\bf 0}_{\tilde{A}}\rangle\langle{\bf 0}_{\tilde{A}}|
\right) \hat{S}_A^\dagger \right] .
\end{equation}
In the following, we assume that the OPO is lossless and 
we often use the mode functions $\Psi_\pm(\Omega)$ 
that are the Fourier transformed expressions of $\Psi_\pm(t)$.

Let us define the complete orthogonal set 
$\{\Psi_\pm(\Omega), \Psi^{(v)}_i(\Omega)\}_i$ in frequency domain 
where $\Psi^{(v)}_i(t)$ corresponds to the mode function 
for the vacuum input. 
Applying it into the annihilation operator $\hat{A}(\Omega)$, 
we have 
\begin{equation}
\label{eq:psi_expansion}
\hat{A}(\Omega) = \Psi_+ (\Omega) \hat{A}_+ 
+ \Psi_- (\Omega) \hat{A}_- 
+ \sum_i \Psi_i^{(v)} (\Omega) \hat{A}_i^{(v)} ,
\end{equation}
where 
\begin{equation}
\label{eq:psi_expansion2}
\hat{A}_i^{(v)} = \frac{1}{2\pi} \int_{-\infty}^{\infty} d\Omega 
\Psi^*_i (\Omega) \hat{A}(\Omega) ,
\end{equation}
and again the superscript $(v)$ means that its initial state is a vacuum.

From the Bogolubov transformation of the OPO 
$\hat{A}^{\rm out} (\Omega) = 
\hat{S}_A^\dagger \hat{A}^{\rm in} (\Omega) \hat{S}_A$, 
which is defined in Eq.~(\ref{eq:Lossless_Bogolubov2}), 
we can describe the input-output relation of the OPO with respect 
to $\hat{A}_\pm$ as 
\begin{eqnarray}
\label{eq:A_pm^out}
\hat{A}_\pm^{\rm out} & = & 
\frac{1}{2\pi} \int_{-\infty}^{\infty} d\Omega 
\Psi_\pm^* (\Omega) \hat{A}^{\rm out} (\Omega) 
\nonumber\\ & = & 
\frac{1}{2\pi} \int_{-\infty}^{\infty} d\Omega 
|\Psi_\pm (\Omega)|^2 \left( 
\bar{\mu}(\Omega) \hat{A}_\pm^{\rm in} 
+ \bar{\nu}(\Omega) \hat{A}_\pm^{{\rm in}\,\dagger}
\right) 
\nonumber\\ &&
+ \frac{1}{2\pi} \int_{-\infty}^{\infty} d\Omega 
\Psi_\pm^* (\Omega) \sum_i \Psi_i^{(v)} (\Omega) 
\nonumber\\ && \times
\left( 
\bar{\mu}(\Omega) \hat{A}_i^{(v)} 
+ \bar{\nu}(\Omega) \hat{A}_i^{(v) \, \dagger}
\right) .
\end{eqnarray}
Note that we have used the relation 
\begin{eqnarray}
\label{eq:pm_orthogonal} 
&&
\frac{1}{2\pi} 
\int_{-\infty}^{\infty} d\Omega 
\bar{\mu} (\Omega) \Psi_+^* (\Omega) \Psi_- (\Omega) 
\nonumber\\ && = 
\frac{1}{2\pi} \int_{-\infty}^{\infty} d\Omega 
\bar{\nu} (\Omega) \Psi_+^* (\Omega) \Psi_- (\Omega) 
= 0 , 
\end{eqnarray}
which implies that the OPO does not couple the modes 
$\Psi_+(\Omega)$ and $\Psi_-(\Omega)$. 
It therefore means that the map $\hat{\mathcal{S}}$ can be decomposed as 
\begin{equation}
\label{eq:S=S+_S-}
\hat{\mathcal{S}} = 
\hat{\mathcal{S}}_+ \otimes \hat{\mathcal{S}}_- .
\end{equation}

\begin{figure}
\begin{center}
\includegraphics[width=0.8\linewidth]{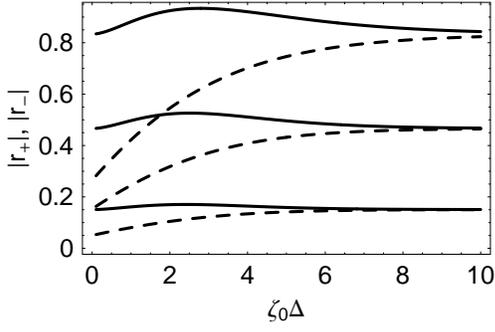}   %
\caption{\label{fig:sq_parameter}
Squeezing parameters $|r_+|$ (solid lines) and $|r_-|$ (dashed lines). 
From the lower to upper curves, $|\epsilon|/\zeta_0 = $ 0.1, 0.3, 0.5, 
respectively. 
}
\end{center}
\end{figure}
\begin{figure}
\begin{center}
\includegraphics[width=0.8\linewidth]{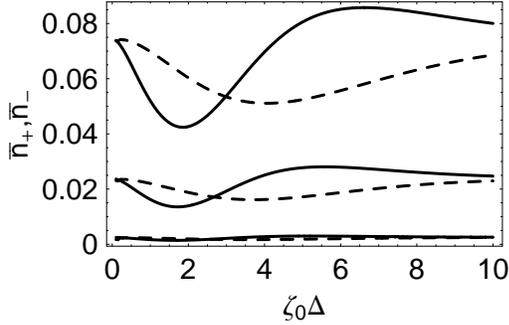}   %
\caption{\label{fig:sq_thermalphoton}
Thermal photons $\bar{n}_+$ (solid lines) and $\bar{n}_-$ (dashed lines).
From the lower to upper curves, $|\epsilon|/\zeta_0 = $ 0.1, 0.3, 0.5, 
respectively. 
}
\end{center}
\end{figure}

The concrete expressions of the maps $\hat{\mathcal{S}}_\pm$ 
are given as follows. 
Since the OPO includes only Gaussian operations, 
$\hat{\mathcal{S}}_\pm$ can be fully characterized by 
the real matrices $S_\pm, Y_\pm$, which describe 
the input-output relation of the covariance matrix as
\begin{equation}
\label{eq:S_pm}
\Gamma_\pm^{\rm out} = S^T_\pm \Gamma_{\rm in} S_\pm + Y_\pm , 
\end{equation}
where $\Gamma_{\rm in}$ and $\Gamma_\pm^{\rm out}$ are 
the covariance matrices of the input and output states, respectively, 
consisting of the variances of the quadratures, e.g. 
\begin{equation}
\label{eq:Gamma_in_appendix}
\Gamma_{\rm in} = \frac{1}{2} \left[
\begin{array}{cc}
\langle \hat{X}_{\rm in}^2 \rangle & 
\langle \hat{X}_{\rm in} \hat{P}_{\rm in} \rangle \\
\langle \hat{P}_{\rm in} \hat{X}_{\rm in} \rangle & 
\langle \hat{P}_{\rm in}^2 \rangle \\ 
\end{array}
\right] .
\end{equation}
From Eq.~(\ref{eq:A_pm^out}), we find 
\begin{eqnarray}
\label{eq:cov_matrix_transformation}
\langle \hat{X}_\pm^{{\rm out} \, 2} \rangle & = & 
G_X^{\pm \, 2} \langle \hat{X}_{\rm in}^2 \rangle 
+ \frac{1}{2} \left( F_X^{\pm} - G_X^{\pm \, 2} \right) ,
\\
\langle \hat{P}_\pm^{{\rm out} \, 2} \rangle & = & 
G_P^{\pm \, 2} \langle \hat{P}_{\rm in}^2 \rangle 
+ \frac{1}{2} \left( F_P^{\pm} - G_P^{\pm \, 2} \right) ,
\\
\langle \hat{X}_\pm^{\rm out} \hat{P}_\pm^{\rm out} \rangle & = & 
G_X^\pm G_P^\pm \langle \hat{X}_{\rm in} \hat{P}_{\rm in} \rangle ,
\end{eqnarray}
where 
\begin{eqnarray}
\label{eq:G}
G_X^\pm & = & 
\frac{1}{2\pi} \int_{-\infty}^{\infty} d\Omega \, 
g(\Omega) |\Psi_\pm (\Omega)|^2 
\nonumber\\ & = & 
\frac{4 \zeta_0^2 \epsilon^2}{\mathcal{N}_\nu (1 \pm I_\Delta)} 
\frac{1}{2\pi} \int_{-\infty}^{\infty} d\Omega
\frac{1 \pm \cos\Delta\Omega}{
(\zeta_+^2 + \Omega^2)^{1/2} (\zeta_-^2 + \Omega^2)^{3/2} } ,
\nonumber\\
\\
G_P^\pm & = & 
\frac{1}{2\pi} \int_{-\infty}^{\infty} d\Omega \, 
g^{-1}(\Omega) |\Psi_\pm (\Omega)|^2 
\nonumber\\ & = & 
\frac{4 \zeta_0^2 \epsilon^2}{\mathcal{N}_\nu (1 \pm I_\Delta)} 
\frac{1}{2\pi} \int_{-\infty}^{\infty} d\Omega
\frac{1 \pm \cos\Delta\Omega}{
(\zeta_+^2 + \Omega^2)^{3/2} (\zeta_-^2 + \Omega^2)^{1/2} } ,
\nonumber\\
\label{eq:G}
F_X^\pm & = & 
\frac{1}{2\pi} \int_{-\infty}^{\infty} d\Omega \, 
g^2(\Omega) |\Psi_\pm (\Omega)|^2 
\nonumber\\ & = & 
\frac{4 \zeta_0^2 \epsilon^2}{\mathcal{N}_\nu (1 \pm I_\Delta)} 
\frac{1}{2\pi} \int_{-\infty}^{\infty} d\Omega \, 
\frac{1 \pm \cos\Delta\Omega}{
(\zeta_+^2 + \Omega^2) (\zeta_-^2 + \Omega^2)^2 } 
\nonumber\\ & = & 
\frac{\zeta_0^2 \epsilon^2 \left\{ 1 \pm (1+\zeta_-\Delta) e^{-\zeta_-\Delta} 
\right\} }{\zeta_-^3 \mathcal{N}_\nu (1 \pm I_\Delta) }
\\
F_P^\pm & = & 
\frac{1}{2\pi} \int_{-\infty}^{\infty} d\Omega \, 
g^{-2}(\Omega) |\Psi_\pm (\Omega)|^2 
\nonumber\\ & = & 
\frac{4 \zeta_0^2 \epsilon^2}{\mathcal{N}_\nu (1 \pm I_\Delta)} 
\frac{1}{2\pi} \int_{-\infty}^{\infty} d\Omega \, 
\frac{1 \pm \cos\Delta\Omega}{
(\zeta_+^2 + \Omega^2)^2 (\zeta_-^2 + \Omega^2) } ,
\nonumber\\ & = & 
\frac{\zeta_0^2 \epsilon^2 \left\{ 1 \pm (1+\zeta_+\Delta) e^{-\zeta_+\Delta} 
\right\} }{\zeta_+^3 \mathcal{N}_\nu (1 \pm I_\Delta) } ,
\end{eqnarray}
and 
\begin{equation}
\label{eq:g(Omega)}
g(\Omega) = \bar{\mu} + \bar{\nu} = \sqrt{
\frac{\zeta_+^2 + \Omega^2}{\zeta_-^2 + \Omega^2} } .
\end{equation}
Then we obtain 
\begin{eqnarray}
\label{eq:S_Y}
S_\pm & = & \left[ 
\begin{array}{cc}
G_X^\pm & 0 \\
0 & G_P^\pm 
\end{array}
\right] ,
\\
Y_\pm & = & \left[
\begin{array}{cc}
F_X^\pm - G_X^{\pm \, 2} & 0 \\
0 & F_P^\pm - G_P^{\pm \, 2}
\end{array}
\right] .
\end{eqnarray}
The above OPO process includes squeezing operation 
and the coupling with a squeezed thermal environment. 
The effective squeezing parameters $r_\pm$ can be defined as
\begin{equation}
\label{eq:r}
r_\pm \equiv -
\frac{1}{2} \ln \frac{G_X^\pm}{G_P^\pm} ,
\end{equation}
and how the process is deviated from a unitary squeezing 
is roughly estimated by looking at 
the numbers of the thermal photons $\bar{n}_\pm$ 
\begin{equation}
\label{eq:nbar}
\bar{n}_\pm \equiv 
G_X^\pm G_P^\pm -1.
\end{equation}
Figs.~\ref{fig:sq_parameter} and \ref{fig:sq_thermalphoton} 
plot $r_\pm$ and $\bar{n}_\pm$, respectively, which clearly 
show that the OPO process with respect to each of $\Psi_\pm$ 
can almost be regarded as a single-mode unitary squeezing.

\section{Wigner function of $\hat{\rho}_+$}

The Wigner function of 
$\hat{\rho}_+$ in Eq.~(\ref{eq:final_state_+}), where 
$\hat{\mathcal{S}}_+$ is applied on the non-Gaussian state 
$C_\phi |\phi\rangle\langle\phi| + C_v |0\rangle\langle0|$, 
is derived from its characteristic function. 
The characteristic function $\chi_+ (u,v)$ is calculable 
with the help of the Bogolubov transformation 
in Eq.~(\ref{eq:A_pm^out}) as 
\begin{eqnarray}
\label{eq:char_func}
\chi_+ (u,v) & = & 
{\rm Tr}\left[ \hat{\rho}_+ 
e^{i(u \hat{X}_+ + v \hat{P}_+)} \right] 
\nonumber\\ & = & 
{\rm Tr}\left[ |\rho_{cw}\rangle\langle\rho_{cw}| 
e^{i(u \hat{X}_+ + v \hat{P}_+)} \right] 
\nonumber\\ & = & 
\langle\rho_{+-}| \langle{\bf 0}_{\tilde{A}}|
\left( \hat{S}_A^\dagger 
e^{i(u \hat{X}_+ + v \hat{P}_+)} \hat{S}_A \right)
|\rho_{+-}\rangle |{\bf 0}_{\tilde{A}}\rangle 
\nonumber\\ & = & 
C_\phi \chi_\phi \left( G_X^+ u , G_P^+ v \right) 
\nonumber\\ && \times
\chi_0 \left( \sqrt{F_X^+ - G_X^{+ \, 2}} u , 
\sqrt{F_P^+ - G_P^{+ \, 2}} v \right) 
\nonumber\\ & & 
+ C_v \chi_0 \left( \sqrt{F_X^+} u , \sqrt{F_P^+} v \right) , 
\end{eqnarray}
where 
\begin{eqnarray}
\label{eq:char_func_phi}
\chi_\phi (u,v) & = & \left\{ 1 
- \frac{c_0 c_2}{\sqrt{2}} (u^2-v^2) + \frac{1}{4}
(u^2+v^2)^2 \right\} 
\nonumber\\ && \times
\exp\left[ -\frac{1}{4}(u^2 + v^2) \right] , 
\\
\label{eq:char_func_0}
\chi_0 (u,v) & = & 
\exp\left[ -\frac{1}{4}(u^2 + v^2) \right] .
\end{eqnarray}
Then its Fourier transformation gives the Wigner function 
\begin{widetext}
\begin{eqnarray}
\label{eq:Wigner_+}
W_+ (x,p) & = & \frac{1}{\pi\sqrt{F_X^+ F_P^+}} 
\left[ 1 + C_\phi \left\{ -\frac{3 c_2^2}{2} \left( 
\frac{G_X^{+ \, 4}}{F_X^{+ \, 2}} + \frac{G_P^{+ \, 4}}{F_P^{+ \, 2}} 
\right) 
+ \frac{G_X^{+ \, 2}}{F_X^{+ \, 2}} \left( 
\sqrt{2}c_2 \left( c_0 + \sqrt{2}c_2 \right) 
- \frac{3 c_2^2 G_X^{+ \, 2}}{F_X^{+}} \right) 
(2x^2 - F_X^+) 
\right. \right.
\nonumber\\ && 
- \frac{G_P^{+ \, 2}}{F_P^{+ \, 2}} \left( 
\sqrt{2}c_2 \left( c_0 - \sqrt{2}c_2 \right) 
+ \frac{3 c_2^2 G_P^{+ \, 2}}{F_P^{+}} \right)
(2p^2 - F_P^+) 
+ \frac{c_2^2 G_X^{+ \, 2}G_P^{+ \, 2}}{F_X^{+ \, 2}F_P^{+ \, 2}} 
(2x^2-F_X^+)(2p^2-F_P^+) 
\nonumber\\ && \left. \left. 
+ 2c_2^2 \left( \frac{G_X^{+ \, 4}}{F_X^{+ \, 4}} x^4 
+ \frac{G_P^{+ \, 4}}{F_P^{+ \, 4}} p^4 \right) 
\right\} \right]
\exp\left[ 
- \frac{x^2}{F_X^+} - \frac{p^2}{F_P^+} \right] .
\end{eqnarray}
\end{widetext}

\acknowledgements

This work was supported by a MEXT Grant-in-Aid for Scientific 
Research (B) 19340115, and a MEXT Grant-in-Aid for Young Scientists (B) 
19740253.

\end{document}